# Mott phase in a van der Waals transition-metal halide at single layer limit

Lang Peng[1§], Jianzhou Zhao[2§], Gui-Yuan Hua[1], Min-Cai[1], Hui-Nan Xia[1], Yuan Yuan[1], Wen-Hao Zhang[1], Gang Xu[1#], Ling-Xiao Zhao[1], Zeng-Wei Zhu[1], Tao Xiang[3], Ying-Shuang Fu[1*]

1. School of Physics and Wuhan National High Magnetic Field Center, Huazhong University of Science and Technology, Wuhan 430074, China

2. Sichuan Co-Innovation Center for New Energetic Materials, Southwest University of Science and Technology, Mianyang 621010, China

3. Institute of Physics, Chinese Academy of Sciences, Beijing 100080, China

Email: [#]gangxu@hust.edu.cn,  [*]yfu@hust.edu.cn

[§] These authors contribute equally to this work.

**Two-dimensional materials offer opportunities for unravelling unprecedented ordered states at single layer limit. Among such ordered states, Mott phase is rarely explored. Here, we report the Mott phase in van der Waals chromium (II) iodide ($CrI_2$) films. High quality $CrI_2$ films with atomically flat surface and macro size are grown on graphitized 6H-SiC(0001) substrate by molecular beam epitaxy. By *in situ* low temperature scanning tunneling microscopy and spectroscopy (STM/STS), we reveal that the film has a band gap as large as ~3.2 eV, which is nearly thickness independent. Density functional plus dynamic mean field theory calculations suggest that $CrI_2$ films may be a strong Mott insulator with a ferromagnetically ordered ground state. The Mott phase is corroborated by the spectral band splitting, that is consistent with the extended Hubbard model, and gap reduction at charge dopants. Our study provides a platform for studying correlated electron states at single layer limit.**

In low-dimensional electronic systems, correlations among electrons are enhanced due to the quantum confinement effect, which favors the formation of long-range collective ordered states[1,2] that are unprecedented according to the Mermin-Wagner theorem[3]. Recent advances in the studies of single-layers of two-dimensional (2D) materials have shown success in the discovery of magnetic order[4,5], charge density waves[6,7], as well as superconductivity[8,9]. The long-range order in single layer limit has delivered surprising quantum behaviors that are distinct from their bulk counterpart, such as magnetic field enhanced magnetism[5], enhanced charge density wave order[7], and Ising superconductivity[8]. Moreover, the ordered states in 2D materials are readily accessible for manipulations with external means[10-12] such as electric field, optical stimuli, etc., which opens a new paradigm for their study and applications.

In Mott insulators, strong electron-electron Coulomb repulsion overwhelms the kinetic hopping energy, which induces gap opening in an otherwise metallic band[13,14]. The Mott physics not only constitutes an intriguing metal-insulator transition mechanism beyond the conventional band theory, but also is relevant to exotic physics of high temperature superconductivity and colossal magnetoresistance. Particularly, the 2D Mott-Hubbard model is widely studied as the basis for generating high Tc superconductivity upon charge doping[15]. While the parent state of cuprate superconductors is a Mott insulator that is considered as 2D identity, there still exists interlayer coupling that prohibits an unambiguous attribution of the property from single layers[15]. Moreover, in Mott systems with John-Teller distortion, the Hubbard bands can split under the cooperative effect of John-Teller distortion and Coulomb repulsion[16]. The resulting Hubbard bands are endowed with distinct orbital characters. This constrains orbital selections for optical excitation, creating dark excitons that aid the development of excitonic insulator phase[17]. In the monolayer limit, the carrier doping of Mott insulators, a crucial parameter for tuning the Coulomb energy, can be controlled with external electrical gating, that is devoid of structural or chemical disorder. To these ends, the exploration of a monolayer Mott insulator with John-Teller effect is highly desirable for in-depth controlled study of rich exotic correlated states, and for building functional devices. Existing experimental single layer Mott systems involve the flat band in magic angle bilayer graphene[18], the surface reconstruction of Sn on Si(111) or Ge(111)[19,20], and single layer 1T-$NbSe_2$.[21-23] However, the Mott mechanism of magic angle graphene is challenged with alternative explanations[24], and the Mott phase in Sn-reconstructed Si(111) or Ge(111) surface is essentially quasi-2D which is stabilized by the substrate. Furthermore, none of the above system exhibits $d$-orbital degeneracy lifting.

$CrI_2$, as a layered van der Waals crystal, has its Cr ion centered around a John-Teller distorted idiom octahedron. The strong Coulomb energy in $d$ orbitals of Cr makes it a promising system for the realization of a single layer Mott insulator with degeneracy lifted Hubbard bands. While this crystal has been synthesized in bulk form, its thin film layers have never been achieved so far. Here, we report the growth of $CrI_2$ films with varying film thickness down to the monolayer limit and identify its Mott insulator phase with STS and theoretical calculations, which features a thickness independent large band gap, characteristic band splitting and gap reduction at dopants.

The CrI$_2$ films were grown on a graphene covered 6H-SiC(0001) substrate with molecular beam epitaxy, whose details are in SI-Note 1. The STM measurements[25] were performed at 4.2 K if not specified exclusively. The STS were performed with a lock-in bias modulation of 14.14 mV (rms) at 829 Hz.

CrI$_2$ is a polymorph of the celebrated ferromagnetic insulator CrI$_3$ [4]. It is a layered van der Waals crystal with monoclinic structure belonging to the space group of C2/m [26]. Its each layer consists of one chromium layer sandwiched by two iodine layers. The top iodine layer forms an isosceles triangle lattice with apex angle of 55° [Figure 1(a) and (b)]. The chromium atom is situated at the center of a distorted iodine octahedron. Figure 1(c) shows a typical STM topographic image of the as-grown thin film. The coverage of the film is about one monolayer (1L), and the step edges on the film are inherited from those of the substrate. There exist small residual areas of 2 L film, which measures the monolayer height as ~6.9 Å [Figure 1(c), inset]. Further growth shows that the film adopts the layer-by-layer mode. Atomic resolution and associated fast Fourier transform [Figure 1(e)] (FFT) of the film [Figure 1(d)] reveal an isosceles triangular lattice with the in-plane lattice constants measured as $a_1$ = 3.88 Å, $a_2$ = 4.23 Å and $a_3$ = 4.18 Å. There are six additional satellite peaks surrounding the center zone as well as each Bragg peaks, which stem from a 6×6 reconstruction between the graphene/SiC(0001) interface, and disappear in film surfaces of higher layers [SI-Figure 1(d-f)]. Since the isosceles triangle lattice is close to a regular one, we have excluded the possibility of inaccuracies in STM topographic acquisition via imaging a twin boundary of the film, where the isosceles triangle lattices of the two domains express strict mirror symmetry to each other [SI-Figure 1(b)]. The measured in-plane lattice constants and the monolayer height unambiguously demonstrate the film is of $CrI_2$, instead of its polymorph $CrI_3$, which has a regular triangular lattice[4].

Next, we characterize the electronic properties of the films. Tunneling spectra, which is proportional to the local density of states, of the monolayer film displays a large band gap and two prominent peaks at –2.6 eV and 1.23 eV, respectively [Figure 2(a)]. The conductance intensity of the 1.23 eV peak drops to zero density at higher energy, resulting in a narrow peak width of 0.9 eV. This implies a large electron mass of the band. Indeed, no standing wave patterns next to impurities or step edges are observed around the associated energy. From the logarithmic scale of the tunneling conductance [Figure 2(b)], the band edges can be clearly resolved, giving a gap size of 3.2 eV. For thicker films from 2$^{nd}$ to 6$^{th}$ layers, the gap size stays the same [Figure 2(c) and (d)], demonstrating the interlayer coupling is negligible. This is distinct from many layered 2D semiconductors, where the band gap size increases with decreasing film thicknesses due to quantum confinement effect[27]. The gap was also measured at 77 K for the films with all different thicknesses, and no difference in spectral shape was detected. The negligible interlayer coupling in conjunction with the narrow peak width and the associated high peak intensity imply electrons in the bands may subject to correlation effect.

To understand the electronic structure of the films, we perform first-principles calculations for the free-standing single-layer $CrI_2$ based on density functional theory (DFT) plus Coulomb interaction U (DFT+U), and density functional theory plus dynamical mean-field theory (DFT+DMFT) methods[28,29] [SI-Note 2]. Spin-orbit coupling effect is negligible [SI-Figure 2] and thus excluded in all calculations. The nonmagnetic (NM) calculations without U show that four 3$d$ electrons of Cr occupy on six nearly degenerated $t_{2g}$ orbitals, giving a metal state of NM CrI$_2$ [Figure 3(a)]. This is obviously contradictory to the large gap observed experimentally. Subsequently, we mainly focus on the magnetic calculations, because stable magnetic state usually accompanies the Cr-based compounds. Five magnetic structures [SI-Figure 3] are considered in our calculations. When U is not included, all the magnetic structures give small insulating gap (~0.1 eV), and a frustrated antiferromagnetic (AFM) configuration, named AFM4 in SI-Figure 3(f), is the most stable. We plot the ferromagnetic (FM) band structures as an example in Figure 3(b) to illustrate the electron occupation in the magnetic states. Namely, for each Cr$^{2+}$ ion, three spin-up $t_{2g}$ orbitals and one spin-up $e_g$ orbital are fully occupied, leaving the other spin-up $e_g$ orbital totally unoccupied with a well separated gap of 0.1 eV due to the John-Teller distortion. All five spin-down orbitals are empty and much higher in energy than the unoccupied spin-up $e_g$ orbital, because of the large Hund's rule coupling in Cr$^{2+}$.

Since the calculated band gap is too small to compare with the experiments, we carry out the DFT+U calculations, and study the evolution of the total energy of the magnetic states and their band gaps with U. The ground state changes from the AFM4 to the FM phase when U is larger than 5 eV [Figure 3(c)]. As the band gaps for all five magnetic phases are nearly the same size, we just plot the results of FM phase in Figure 3(c). Remarkably, a nominal U of 12 eV is needed to open an experimentally comparable gap of 3.18 eV, whose band structures are shown in Figure 3(d), indicating that the large band gap of CrI$_2$ has a Mott origin. Similar calculations on bulk CrI$_2$ deliver identical results as the monolayer. This demonstrates the interlayer coupling is negligible, which is consistent with the experiment. However, such U is abnormally large for Cr-compounds[30,31]. One possible reason is that the effective U added to the correlated electrons in the DFT+U method could be severely screened by the other electrons[32,33].

To more reasonably describe effective U, we performed DFT+DMFT calculations on the single-layer $CrI_2$. We mainly focus on its FM state, which has shown to be the ground state in the DFT+U (U > 5 eV) calculations. Its band gap increases rapidly with U from 1 eV to 4 eV [SI-Figure 4]. A gap of ~3 eV is obtained with U = 4 eV, which is a typical value for Cr-compounds, demonstrating the correlation effect in $CrI_2$ is reasonably captured. The corresponding momentum resolved spectral [Figure 3(e)] and projected density of states [SI-Figure 5] show that the highest valence bands (HVB) and the lowest conduction bands (LCB) are both contributed by the spin-up *3d* orbitals of $Cr^{2+}$ ions with $d_{z^2}$ and $d_{x^2-y^2}$ characters, namely, the lower Hubbard bands (LHBs). Thereafter, the HVB and the LCB are denoted as LHB($d_{z^2}$) and LHB($d_{x^2-y^2}$), respectively. The upper Hubbard bands (UHBs) are much higher in energy and are beyond the spectroscopic range of our measurement. Since $d_{z^2}$ is more localized compared to $d_{x^2-y^2}$, its effective Coulomb energy is larger. This renders the energy splitting between LHB and UHB, which is related to the onsite Coulomb energy, is larger for $d_{z^2}$ than that of $d_{x^2-y^2}$ orbital. The distinct orbital characters of the $CrI_2$ bands are a result of cooperative Coulomb repulsion and Jahn-Teller distortion of the $I^-$ octahedra, which is analogous to the perovskites $KCrF_3$[34] and $LaMnO_3$.[35]

Finally, we also calculated the electronic structure of the paramagnetic (PM) phase for single-layer CrI$_2$ by the DFT+DMFT method. The calculated density of states [SI-Figure 4(b)] and the momentum resolved spectral [Figure 3(f)] both show an insulating gap of ~ 3 eV with U = 4 eV. These demonstrate that the large gap exist at 300 K already, ruling out the possibility of Slater transition. For Slater insulator, the insulating gap only exists in the magnetic phase, while their PM phase is metallic[36].

The CrI$_2$ films have abundant intrinsic defects, which appear as depression spots under positive bias [SI-Figure 6 (a)]. Atomic resolution reveals its origin as local lattice distortion, involving three surface iodine atoms at the defect site [SI-Figure 6 (b)]. Line spectra [Figure 4(b)] show rigid band shift towards higher energies, when approaching the defect [Figure 4(a)]. This indicates that the defect acts as hole-dopant. Intriguingly, there appear two sets of peaks for LHB($d_{z^2}$) and LHB($d_{x^2-y^2}$) at the defect center, whose intensities are strongly localized within ~2 atomic lattices. The band gap at the defect site becomes smaller by ~0.37 eV. This behavior is reminiscent of the spectroscopic features for charge dopants in Mott insulators[37,38]. Around the defect, charge doping emphasizes the importance of nonlocal interactions, and calls for the extended Hubbard model, in which both the on-site Coulomb interaction *U*, and the nearest neighbor Coulomb repulsion *V* should be considered. While *U* determines the energy difference between the upper and lower Hubbard bands, the *V*, which is overlooked in the Hubbard model, manifests itself at the charge dopants. In such case, charge dopants invoke charged excitons accessible to electron tunneling spectroscopy[39], which induces satellite peaks below the Hubbard bands with splitting energy *V*.[40] Moreover, since the Hubbard bands arise from multi-orbitals in CrI$_2$ film, corresponding *V* for the different orbitals are inequivalent. From the spectral splitting in Figure 4(c), values of *V* are determined as 0.46 eV and 0.26 eV for $d_{z^2}$ and $d_{x^2-y^2}$ orbitals, respectively.

The Mott phase in CrI$_2$ is further examined by doping electrons, which expects Mott gap reduction due to locally decreased Coulomb repulsion by the charge dopants. Na and Zn atoms, which lack unfilled *d* electrons and merely act as charge dopants, are deposited onto the CrI$_2$ surface at ~30 K, forming clusters. The gap size on clusters of the both species becomes significantly smaller and recovers right after leaving the clusters [SI-Figure 7]. Moreover, the gap size decreases with increasing cluster size. This excludes the possibility of a Coulomb blockade gap, whose gap size would show opposite trend with cluster size. This observation conforms to the attribution of CrI$_2$ to Mott insulator.

In summary, CrI$_2$ films down to the single-layer limit are prepared in a layer-by-layer growth mode. The films have a large band gap of ~3.2 eV that is nearly independent of the film thickness. DFT+U and DMFT calculations suggest that the large band gap originates from a Mott phase with out-of-plane ferro-magnetization. The correlated states in CrI$_2$ films envision in-depth further studies, including characterizing their magnetic properties, tuning their properties with external parameters such as electric gating, as well as developing functional devices. Moreover, the orbital character of the Hubbard bands implies constrained selection rules for optical absorption, which provides a candidate system for exploring dark excitons and half excitonic insulator[17].

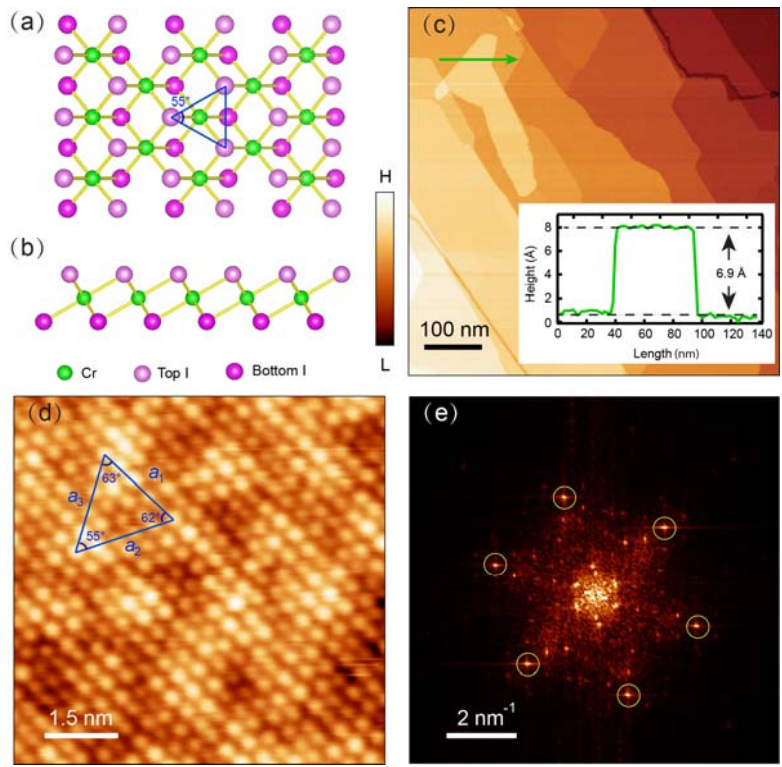

**Figure 1. Crystal structure of CrI$_2$ and its morphology.** (a, b) Top view (a) and side view (b) of the crystal structure of monolayer CrI$_2$. (c) Large scale STM topographic image ($V_s$ = 3.0 V, $I_t$ = 6 pA) of CrI$_2$ film. Line profile (inset) along the green line shows the monolayer step height. (d) STM image ($V_s$ = 0.6 V, $I_t$ = 100 pA) showing atomic resolution of 1L CrI$_2$. The top layer iodine atoms form an isosceles triangle lattice. (e) FFT image of (d). The yellow circles highlight the Bragg peaks of the iodine lattice.

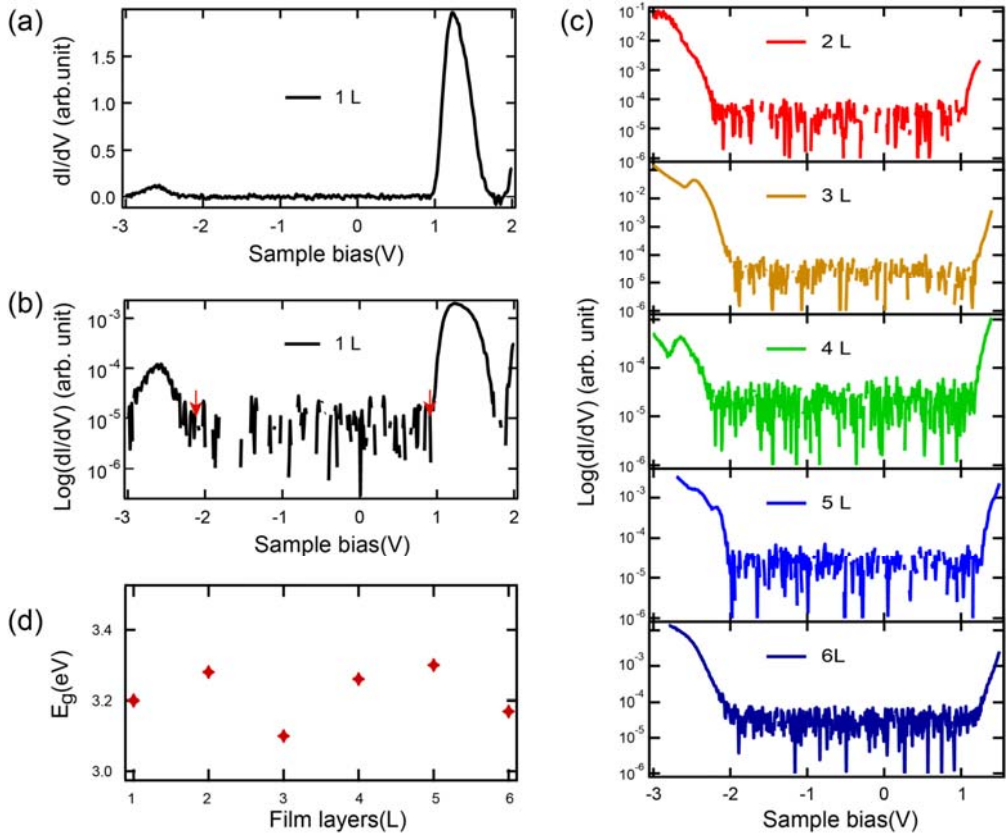

**Figure 2. Tunneling spectra of CrI$_2$ films.** (a,b) Tunneling spectroscopy of 1L CrI$_2$ in linear (a) and logarithmic scale (b) of its conductance, respectively. The red arrows in (b) indicate the band edges. (c) Tunneling spectra of multilayer CrI$_2$ films from 2L to 6L in logarithmic scale of their conductance. (d) Energy gap of the CrI$_2$ with different film thicknesses measured from (b) and (c). Set point conditions: $V_s$ = 2.0 V, $I_t$ = 50 pA for 1L; $V_s$ = 1.25 V, $I_t$ = 10 pA for 2L and 3L; $V_s$ = 1.4 V, $I_t$ = 10 pA for 4L; $V_s$ = 1.4 V, $I_t$ = 5 pA for 5L; $V_s$ = 1.55 V, $I_t$ = 20 pA for 6L.

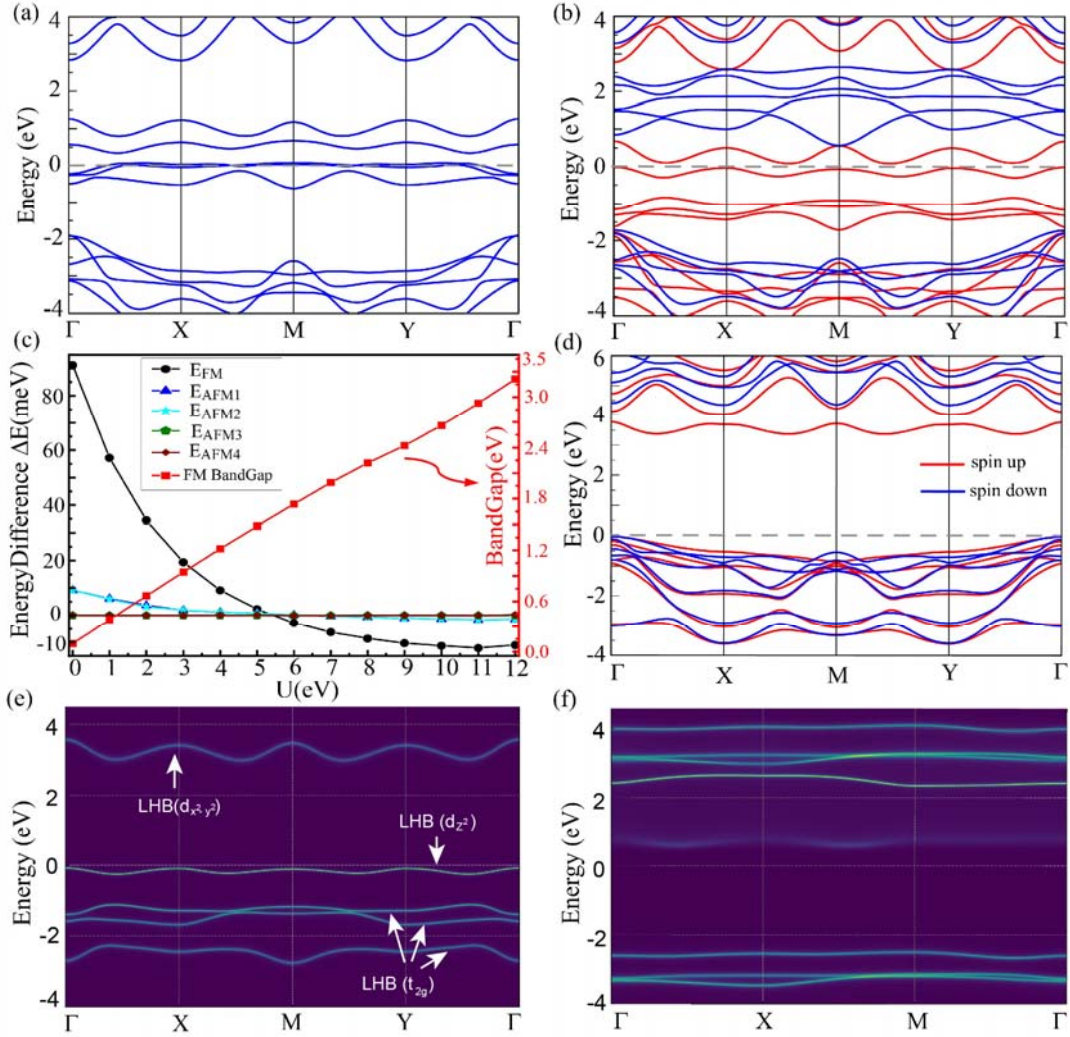

**Figure 3. Calculations of monolayer CrI$_2$.** (a, b) Band structures of single-layer CrI$_2$ in NM (a) and FM state (b) calculated by DFT. (c) Calculated energy difference ΔE (left axis) and band gap of FM phase (right axis) versus U with DFT+U method. Energy difference is defined as ΔE=E - E$_{AFM4}$, where E and E$_{AFM4}$ denote the total energy of different magnetic configurations and the AFM4 configuration. (d) Band structure of CrI$_2$ in FM phase calculated by DFT+U method with U = 12 eV. In (b) and (d), the red (blue) bands denote the spin-up (spin-down) channel. (e, f) Momentum resolved spectral of single-layer CrI$_2$ by DFT+DMFT method with U = 4.0 eV and J=0.8 eV in FM phase (e) and PM phase (f). The orbital characters of the bands are marked in (e).

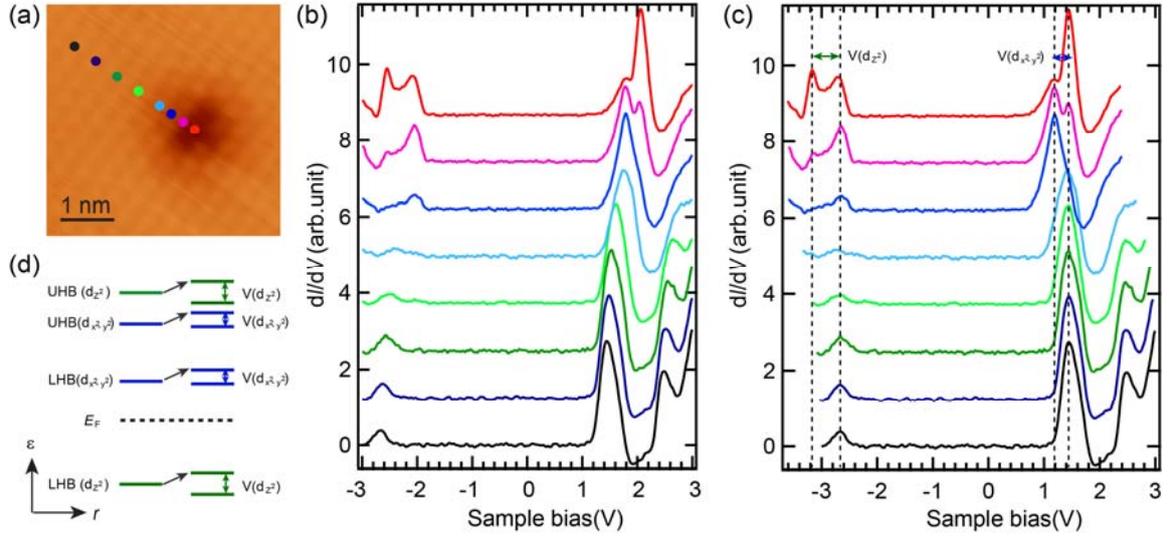

**Figure 4. Morphology and tunneling spectra of defects.** (a) STM image ($V_s$ = 1.75 V, $I_t$ = 100 pA) of a defect on 2$^{nd}$ layer CrI2. (b) Tunneling spectra measured at different distances relative to the defect, where the spectroscopic locations are indicated in (a). Spectra set points: $V_s$ = 3.0 V, $I_t$ = 50 pA. The spectral curves are offset vertically for clarity. (c) Spectra of (b) after being aligned to the right peak of LCB, corresponding to the LHB($d_{x^2-y^2}$). The dashed lines mark the energy splitting for the different Hubbard bands. (d) Schematics showing the energy of Hubbard bands near the Fermi level from $d_{z^2}$ (green lines) and $d_{x^2-y^2}$ (blue lines) orbitals, respectively. At the *p*-type defect site, nearest neighbor Coulomb interaction ($V$) induces satellite peaks with splitting energy $V$ below the Hubbard bands. Note that the splitting energy for different orbitals are inequivalent.

**Acknowledgement:**

We thank S.W. Wu and J.T. Lü for discussions. This work is funded by the National Key Research and Development Program of China (Grant No. 2017YFA0403501, 2018YFA0307000, 2016YFA0401003), the National Science Foundation of China (Grants No. 11522431, No. 11474112, No. 11774105, No. 11874022, No. 21873033).